\documentclass[journal,twoside]{IEEEtran}
\usepackage{amssymb}
\usepackage{amsthm}
\usepackage{multirow}
\usepackage{inputenc}
\usepackage{enumerate} 
\usepackage{textcomp}
\usepackage{listings}
\usepackage{array}
\usepackage[switch,mathlines]{lineno}
\usepackage{soul}
\usepackage{xfrac}
\usepackage{graphicx}
\usepackage{cite}
\usepackage[cmex10]{amsmath}
\usepackage{algorithm}
\usepackage{xcolor}
\usepackage{colortbl}

\usepackage{cancel}

\usepackage{algorithm,algpseudocode}
\algnewcommand{\Inputs}[1]{%
  \State \textbf{Inputs:}
  \Statex \hspace*{\algorithmicindent}\parbox[t]{.8\linewidth}{\raggedright #1}
}
\algnewcommand{\Initialize}[1]{%
  \State \textbf{Initialize:}
  \Statex \hspace*{\algorithmicindent}\parbox[t]{.8\linewidth}{\raggedright #1}
}

\makeatletter
\def\footnoterule{\kern-3\p@
  \hrule \@width 3.3in \kern 2.6\p@} 
\makeatother

\usepackage{algpseudocode}

\usepackage{stackengine}

\usepackage{algpseudocode}
\usepackage{CJK}
\usepackage[cmex10]{amsmath}
\usepackage{bm}
\usepackage{color}
\usepackage{amsmath,amsthm,amssymb,amsfonts}
\usepackage{flushend}
\usepackage{float}

\newtheorem{remark}{Remark}

\usepackage{arydshln} 
\usepackage{hyperref}
\hypersetup{
     colorlinks   = true,
     linkcolor    = blue,
     citecolor    = red,
     urlcolor     = blue
}
\makeatletter
\newcommand*{\transpose}{%
  {\mathpalette\@transpose{}}%
}
\newcommand*{\@transpose}[2]{%
  \raisebox{\depth}{$\m@th#1\intercal$}%
}
\makeatother

\usepackage{soul}
\usepackage{tikz}
\usepackage{booktabs}
\usepackage{tabularx}

\usepackage[caption=false,font=footnotesize]{subfig}
\ifCLASSINFOpdf
\else
\fi

\begin{document}
\renewcommand{\ttdefault}{cmtt}
\bstctlcite{IEEEexample:BSTcontrol}

\title{Alleviating CoD in Renewable Energy Profile Clustering Using an Optical Quantum Computer}



\author{{
Chengjun Liu,~\IEEEmembership{Student Member},
Yijun~Xu,~\IEEEmembership{Senior Member}, 
Wei Gu,~\IEEEmembership{Senior Member},
Bo Sun,~\IEEEmembership{Student Member},\\
Kai Wen,
Shuai Lu,~\IEEEmembership{Member}, 
and Lamine Mili,~\IEEEmembership{Life Fellow}
}

}

\markboth{CSEE Journal of Power and Energy Systems}%
{Liu \MakeLowercase{\textit{\textit{et al.}}}: Alleviating CoD in Renewable Energy Profile Clustering Using an Optical Quantum Computer}
\maketitle
\begin{abstract}
The traditional clustering problem of renewable energy profiles is typically formulated as a combinatorial optimization that suffers from the Curse of Dimensionality (CoD) on classical computers. To address this issue, this paper first proposed a kernel-based quantum clustering method. More specifically, the kernel-based similarity between profiles with minimal intra-group distance is encoded into the ground-state of the Hamiltonian in the form of an Ising model. 
Then, this NP-hard problem can be reformulated into a Quadratic Unconstrained Binary Optimization (QUBO), which a Coherent Ising Machine (CIM) can naturally solve with significant improvement over classical computers. The test results from a real optical quantum computer verify the validity of the proposed method. It also demonstrates its ability to address CoD in an NP-hard clustering problem. 
\end{abstract}
\vspace{-0.2cm}

\IEEEpeerreviewmaketitle
 \vspace{-0.4cm}
\section{Introduction}
\IEEEPARstart{C}{lustering} technique
serves as a fundamental tool in data analysis\cite{9420350} and aggregation control\cite{9171660} for power systems. However, it is typically formulated as a combinatorial optimization problem that inevitably suffers CoD restriction
by classical computers based on integrated circuits.

Fortunately, the development of quantum computing makes it possible to solve the CoD\cite{morstyn2024opportunities} as demonstrated by several researches. For example, Xie \emph{et al.} proposes a quantum-based algorithm to solve a coordinated post-disaster restoration model for urban distribution systems\cite{FU2023129314}. Zheng \emph{et al.} construct a quantum algorithm for solving unit commitment problems in Grover's framework \cite{10381746}. In addition, a quantum annealing method is advocated in \cite{9863874} to solve combinatorial optimal power flow. 
Similarly, the clustering problem of renewable energy profiles \cite{10322714},\cite{10552141} that we are addressing in this paper is such a combinatorial optimization problem that naturally meets the CoD restricted by the classical computer.

To address this issue, this paper, for the first time, proposes an optical quantum computer-based method to overcome CoD in profile clustering optimization.
More specifically, we first illustrate the connection between the Ising model, the CIM, and the typical QUBO model.  
Using it, we further formulate a kernel-based clustering method into a QUBO problem that matches the Hamiltonian of an Ising model. 
Interestingly, this can be directly solved by a CIM \cite{PRA}. 
Experiments from a real 400 qubits CIM from QBoson \cite{bose} reveal the excellent performance of the proposed method. 

\vspace{-0.2cm}
\section{Problem Statement}
The renewable energy profile represents the power output recorded at each measurement point throughout the day. In this paper, photovoltaic (PV) profiles are selected as a representative example.  Assume that there are $N$ profiles to be clustered into $G$ groups to minimize the intra-group distance. 
This leads to a discrete optimization objective, $H_\text{obj}$, expressed as
\begin{equation}
H_\text{obj} = \frac{1}{2} \underset{g}{\sum} \underset{y_{i}=y_{j}}{\sum}d_{i,j},
\end{equation}
where $g \in \{1,2,…,G\}$ is a group index,  $y_{i},\enspace y_{j} \in \{1, 2,..., G\}$ $\left(i,\enspace j \in \{1, 2 , \ldots , N\}\right)$ is an integer variable that shows to which group profile $i$ belongs, $y_{i}=y_{j}$ means profile $i$ and $j$ are assigned to the same group, and $d_{i, j}$ denotes the Euclidean distance between profile $i$ and $j$ to quantify the similarity of the data. 

In practice, finding the minimum of this objective function is difficult due to the strong non-convexity of the feasible domain. To overcome this difficulty, $y_{i}$ is usually divided into $G$ binary variables $\left[x_{i}^{1}, ...,x_{i}^{g},... x_{i}^{G}\right]$, in which $x_{i}^{g} \in \{0,1\} $ is a binary decision variable. Here, $1$ means that the profile $i$ belongs to the group $g$. In addition, since each profile belongs to a single group, the constraint ${\sum}_{g} x_{i}^{g} = 1$ is needed. These transform the clustering problem into the following form:
\begin{equation}
    \label{CCO}
    \begin{gathered}
        \min_{{x_{i}^{g}}, x_{j}^{g}}\enspace \frac{1}{2}\underset{i}{\sum} \underset{j}{\sum} d_{i, j} \underset{g}{\sum} x_{i}^{g} x_{j}^{g}, \\
       \textrm{s.t}.\enspace \underset{g}{\sum} x_{i}^{g} = 1, \forall i \in \{1, 2 , \ldots , N\}.
    \end{gathered}
\end{equation}
\vspace{-0.2cm}

This is a constrained combinatorial optimization problem that is NP-hard for a classical computer.

\vspace{-0.2cm}
\section{Methodology}
Inspired by the Ising model in physics, we transform the aforementioned PV profile clustering problem into a QUBO model that is applicable to a CIM.
\subsection{From Ising Model to QUBO}

\subsubsection{Ising Model}
\begin{figure}[!htbp]
    \centering
     \includegraphics[width=0.9\linewidth]{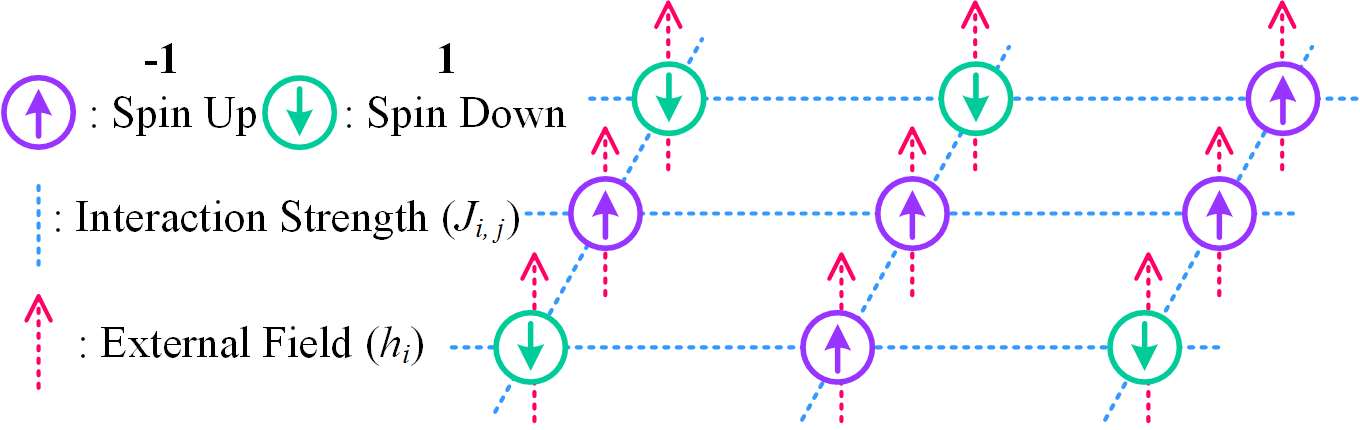}
     \vspace{-0.2cm}
    \caption{Plot for the visualization of 2D Ising model where $s_i = +1$ means the lattice is spin up while $s_i = -1$ means the lattice is spin down.}
    \label{fig1}
\end{figure}

The Ising model originated in physics to describe the phase transition process \cite{brush1967history}. For a lattice of spins that point up or down, each spin is independently affected by the external field, as shown in Fig. ~\ref{fig1}. Here, the energy function (or Hamiltonian), $H_\text{Ising}$, of the whole system is expressed as
\vspace{-0.2cm}
\begin{equation}
\label{Ising}
    H_\text{Ising} = \underset{i \textless j}{\sum} J_{i, j} s_{i} s_{j} + \underset{i}{\sum} h_{i} s_{i},
\end{equation}
where  $s_{i} \in \{-1, +1\}$ is the spin state variable, $h_{i}$ is a constant reflecting the degree of influence from the external field on the $s_{i}$, and $J_{i, j}$ is a coupling coefficient reflecting the interaction strength between $s_{i}$ and $s_{j}$. Note that $J_{i, j} \neq 0$ if $s_{i}$ and $s_{j}$ are neighbors; otherwise, $J_{i, j} = 0$. 

\subsubsection{CIM}
CIM is a specialized quantum computer for solving minimization problems based on the Ising model, consisting of a Degenerate Optical Parametric Oscillator (DOPO) network. In this network, each DOPO represents a spin in the Ising model. These coupled oscillators interact through coherent signal injection, driving the network toward modes that minimize photon decay, following the minimum gain principle \cite{PRA}.

\begin{figure}[!htbp]
    \centering
     \includegraphics[width=0.9\linewidth]{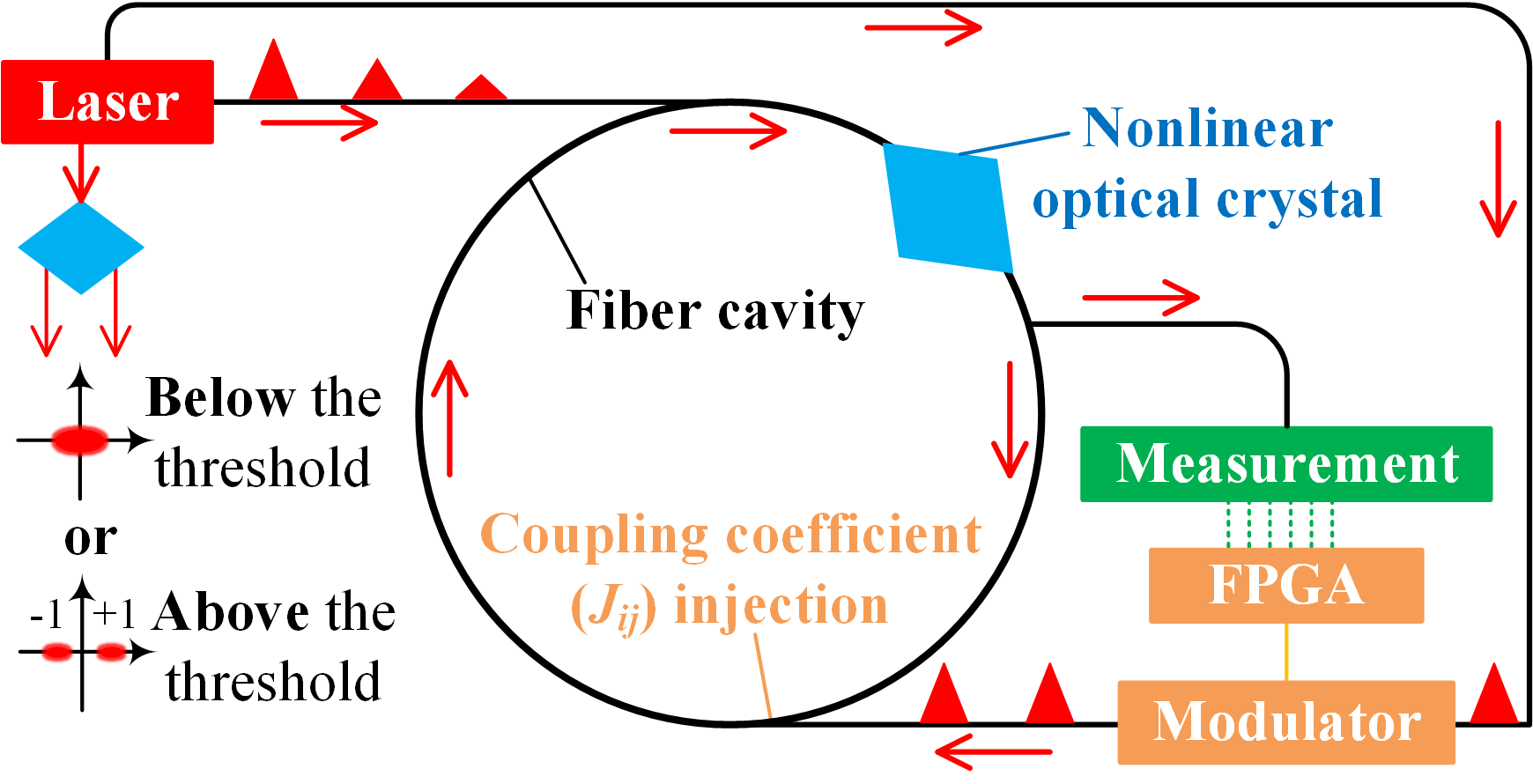}
    \caption{Plot for the basic principle of the CIM. In a DOPO, a photon pulse generated by the laser source passing through a nonlinear optical crystal splits into two pulses. When the pump rate is low, the system is in a \textquotedblleft compressed state\textquotedblright, making the photon's phase indistinguishable. However, once the pump rate exceeds a threshold, it transitions to a \textquotedblleft coherent state\textquotedblright, where each DOPO oscillates in one of two phases ($0$ or $\pi$), corresponding to spin states ($+1$ or $-1$).
A portion of the photon pulse in the fiber cavity is measured and fed into a Field-Programmable Gate Array (FPGA) to calculate the coupling coefficients in the Ising model, then the results are encoded by the modulator onto the feedback photon pulse and re-injected into the fiber cavity as $J_{i,j}$.}
    \label{fig2}
\vspace{-0.3cm}
\end{figure}

As shown in Fig.~\ref{fig2}, the laser source and DOPOs generate qubits while the measurement-feedback part injects the coupling coefficients. For more details, refer to our previous work \cite{PRA}. By mapping the Hamiltonian of the Ising model to the losses in the DOPO network, the network can spontaneously identify the combination of spin states that minimizes this loss, facilitating auto-minimization of the Ising model.

\begin{remark}
The computing time for CIM is almost constant no matter the number of variables, since it is always equal to the time that a photon pulse takes to complete a lap in the fiber cavity. This is the key to a CIM that overcomes the CoD that classical computers suffer from. This phenomenon will be verified in Section IV. 
Another advantage of CIM is that, unlike superconducting quantum computers, it does not require the design of a corresponding variational quantum circuit, mitigating issues associated with circuit depth.
Besides, compared to mainstream quantum annealers such as D-Wave, CIM supports native all-to-all connectivity and continuous parallel updates, offering superior performance in combinatorial opitimiaztaions \cite{PRA}.
\end{remark}

\subsubsection{QUBO in CIM}
Since the CIM needs to download the optimization model from the host computer via FPGA while the storage of the Ising model is not efficient in a classic computer, it is necessary to convert it into a QUBO problem. In general, a QUBO objective function can be represented as
\vspace{-0.2cm}
\begin{equation}
\label{fqubo}
f_\text{QUBO} \left(x\right) = \underset{i}{\sum} \underset{j}{\sum} q_{i, j} x_{i} x_{j} + \underset{i}{\sum} q_{i} x_{i} + C,
\end{equation}
where $x_{i} \in \{0,1\}$ is a binary variable, $q_{i, j}$ and $q_{i}$ are coefficients, and $C$ is a constant. The QUBO problem is isomorphic to the Ising model, and the detailed derivation is provided in the appendix.

Furthermore, since $\left(x_{i}\right) ^2 = x_{i}$,  we have
\begin{equation}
    f_\text{QUBO} \left(x\right) = \bm{x}^{T} \bm{Q} \bm{x},
\end{equation}
where $\bm{x}$ is the variable vector and $\bm{Q}$ is an upper triangular coefficient matrix that can be imported to the CIM.  
\subsection{Combinatorial Kernel PV Profile Clustering Using QUBO}
\subsubsection{Conventional PV Clustering in CIM}
In the PV profile clustering, since \eqref{CCO} is not a solvable QUBO in CIM,  let us take ${\sum}_{g} x_{i}^{g} = 1$ as the penalty to rebuild the QUBO form objective function, $H_\text{QUBO}$,  given by
\begin{equation}  
\label{Hqubo}
H_\text{QUBO} = \underbrace{\frac{1}{2}\underset{i}{\sum} \underset{j}{\sum} d_{i , j} \underset{g}{\sum} x_{i}^{g}  x_{j}^{g}}_{H_\text{obj}} + \underbrace{\underset{i}{\sum} \lambda_{i} \underset{g}{\sum} \left(x_{i}^{g} - 1\right)^{2}}_{H_\text{cons}},
\end{equation}
where $\lambda_{i}$ is the penalty coefficient. As proved in \cite{kumar2018quantum}, $\lambda_{i}$ should satisfy
\begin{equation}
\lambda_{i} \geq \left(N - G\right)  d_{i , j}^\text{max},
\end{equation}
where $d_{i , j}^\text{max}$ is the maximum of $d_{i , j}$ for all profiles.

Let us put the polynomial form given by \eqref{Hqubo} in matrix form. To this end, we transform its objective term as follows:
\begin{equation}
H_\text{obj} = \frac{1}{2}\underset{i}{\sum} \underset{j}{\sum} d_{i  j} \underset{g}{\sum} x_{i}^{g} x_{j}^{g} = \bm{x}^{T}\left(\bm{{D}_\text{u}} \bm\otimes \bm{I} \right) \bm{x},
\end{equation}
where  $\bm{x} = \left[x_{1}^{1}, \ldots, x_{1}^{G}, \ldots, x_{N}^{1}, \ldots, x_{N}^{G}\right]^{T}$, $\bm{{D}}_\text{u} \in \mathbb{R}^{N \times N}$ is an upper triangle matrix of $\bm{D}$, with $d_{i,j}$ as elements, and $\bm{I} \in \mathbb{R}^{G \times G} $ is an identity matrix. Here,  $\bm\otimes$ is the Kronecker product. 
Then, omitting the constant term, the constraint part, $H_\text{cons}$, can be expressed as
\begin{equation}
\label{Hcons}
\begin{split}
H_\text{cons} & = - \underset{i}{\sum} \lambda_{i} \underset{g}{\sum} \left(x_{i}^{g}\right)^{2} + 2 \underset{i}{\sum} \lambda_{i} \underset{g \textless h}{\sum}x_{i}^{g} x_{j}^{h} \\
& = \bm{x}^{T}\left(\left(-\bm{\Lambda \otimes I} +2\bm{ \Lambda \otimes O}\right)\right)\bm{x} \\
& = \bm{x}^{T} \left(\bm{\Lambda \otimes} \left(2\bm{O-I}\right) \right) \bm{x},
\end{split}
\end{equation}
where $\bm{\Lambda} \in \mathbb{R}^{N \times N}$ is a diagonal matrix with $\lambda_{i}$ for the diagonal elements and $\bm{O} \in \mathbb{R}^{G \times G}$ is an upper triangle matrix with $0$ for the diagonal elements and $1$ for the other upper triangular elements.

Thus, the QUBO matrix form given by  \eqref{Hqubo} is expressed as
\begin{equation}
\label{Hqubo2}
H_\text{QUBO} = \bm{x}^{T}\left(\left(\bm{{D}_\text{u} \otimes I}\right)+ \left(\bm{\Lambda \otimes} \left(2\bm{O-I}\right)\right)\right) \bm{x}.
\end{equation}

\subsubsection{Kernel-based PV Clustering in CIM}
Recently, kernel-based method has become popular in clustering due to its ability to better characterize the data structure \cite{camastra2005novel}. 
However, like conventional clustering methods, the kernel clustering method is also affected by CoD.
In this paper, we overcome the CoD of the classical kernel clustering method, by mapping the method to the QUBO model and solving it on the CIM.
To put \eqref{Hqubo2} in the kernel form, let us replace the Euclidean distance, $d_{i,j}$, with a nonlinear kernel-based similarity index, $g_{i,j}$, defined as
\begin{equation}
g_{i , j} = k_{i , j} - \frac{\sum_{c} k_{i , c}}{N} - \frac{\sum_{r} k_{r , j}}{N} - \frac{\sum_{r} \sum_{c} k_{r , c}}{N^{2}},
\end{equation}
where $c \in \{1,2,…,N\}$ is an index for columns and $r \in \{1,2,…,N\}$ is an index for rows. 
$k_{i, j}$ denotes a nonlinear Gaussian kernel function, $k_{i, j} = e^{- \frac{1}{2 \sigma^{2}} d_{i, j}}$.

Now, let us define $\bm{G} \in \mathbb{R}^{N \times N}$ as the matrix with $g_{i,j}$ as elements. The matrix then contains all the similarity information for clustering. Note that the diagonal elements in $\bm{G}$ are non-zeros, which is different from $\bm{D}$. Then, the objective of kernel clustering is given by
\begin{equation}
\begin{split}
H_\text{obj}^\text{kernel} & = - \underset{i}{\sum} \underset{j}{\sum} g_{i , j} \underset{g}{\sum} x_{i}^{g} x_{j}^{g} \\
& = - \underset{i}{\sum} g_{i , i} \underset{g}{\sum} x_{i}^{g} x_{j}^{g} - 2 \underset{i \textless j}{\sum} g_{i , j} \underset{g}{\sum} x_{i}^{g} x_{j}^{g}.
\end{split}
\end{equation}

Its matrix form can be further expressed as
\begin{equation}
H_\text{obj}^\text{kernel}= \bm{x}^{T}\left(\left(\bm{-{G}_\text{diag}}-2\bm{G}_\text{u}\right) \bm\otimes \bm{I} \right) \bm{x},
\end{equation}
where $\bm{{G}_\text{diag}} \in \mathbb{R}^{N \times N}$ is a diagonal matrix with only the diagonal elements in $\bm{G}$. $\bm{{G}_\text{u}} \in \mathbb{R}^{N \times N}$ is an upper triangle matrix with $0$ for the diagonal elements and the upper triangular elements of $\bm{G}$.

Since \eqref{Hcons} remains unchanged, the QUBO model for the kernel clustering method in the matrix form is expressed as
\begin{equation}
\begin{split}
H_\text{QUBO}^\text{kernel} = \bm{x}^{T} & \left(\left(\bm{-{G}_\text{diag}}-2\bm{G}_\text{u}\right) \bm\otimes \bm{I}\right) \\
& + \left(\bm{\Lambda \otimes} \left(2\bm{O-I}\right)\right) \bm{x}.
\end{split}
\end{equation}

This is exactly a QUBO form applicable to the CIM.

\vspace{-0.3cm}
\section{Case Studies and Discussion}
This section presents case studies on a real quantum computer and discusses the limitations of both the proposed method and the quantum hardware.
\subsection{Case Studies}
Now, let us test the proposed method in a $400$ qubits CIM from Beijing QBoson Quantum Technology Co., Ltd \cite{bose}. The PV profiles are obtained from the Yulara Solar System \cite{Australia}, collected at 5-minute intervals and averaged to hourly resolution. 
Comparison algorithms include K-means (MATLAB), K-medoids (MATLAB), Simulated Quantum Annealing (SQA, OpenJij) and Gurobi 11.0 in a classical computer (i5-13400F 2.5GHz, 32G RAM). Here, we set $4$ cases with different variable scales: clustering $50/60/70/80$ profiles into $2/3/4/5$ groups, including $100/180/280/400$ binary variables, respectively. The similarity matrix $\bm{G}$ is normalized to $\left[0, 1\right]$, and the penalty coefficient $\lambda_{i}$ is set to be 10 times the maximum element \cite{kumar2018quantum} in $\bm{G}$, ensuring that all matrix entries are within the acceptable bit-width limits of quantum hardware. The silhouette coefficient proposed by Rousseeuw \cite{rousseeuw1987silhouettes} is employed to verify the effectiveness of the clustering.

\begin{figure}
    \centering
    \includegraphics[scale=1]{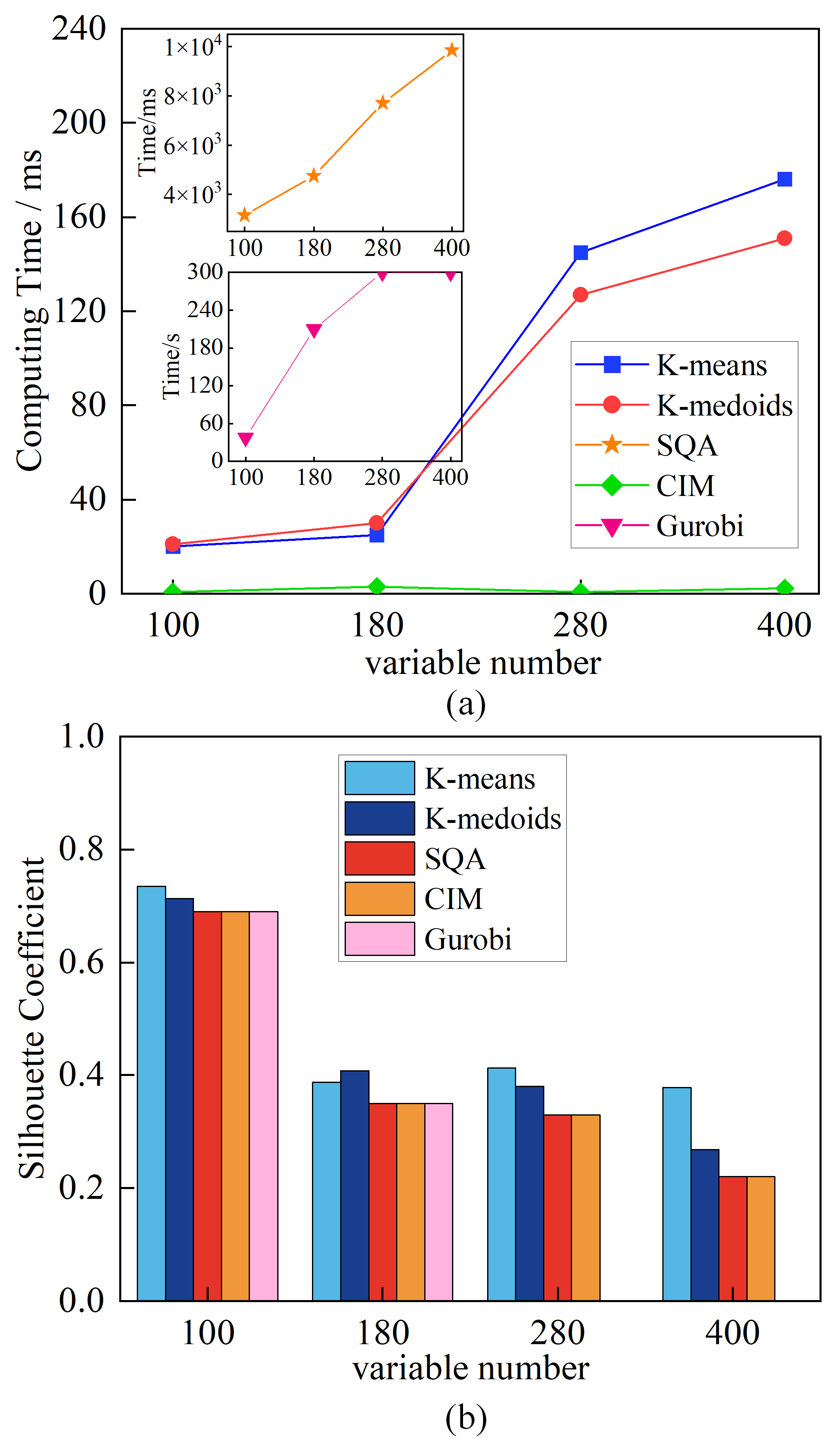}
    \vspace{-0.4cm}
    \caption{Plots for clustering results of different algorithms from the classical computer and CIM under different numbers of variables: $\text{(a)}$ computing time; $\text{(b)}$ silhouette coefficient. The reported computing times for all methods refer exclusively to the core execution time of the algorithm itself, ensuring a fair and consistent basis for comparison across classical and quantum-inspired solvers. For classical algorithms such as K-means, K-medoids, and SQA, this includes all iterative procedures until convergence. For Gurobi, it corresponds to the optimization procedure of solving the QUBO model using a branch-and-bound framework. For CIM, it represents the physical procedure for optimization.}
    \label{fig3}
    \vspace{-0.5cm}
\end{figure}

\begin{figure}
    \centering
    
    \includegraphics[scale=1]{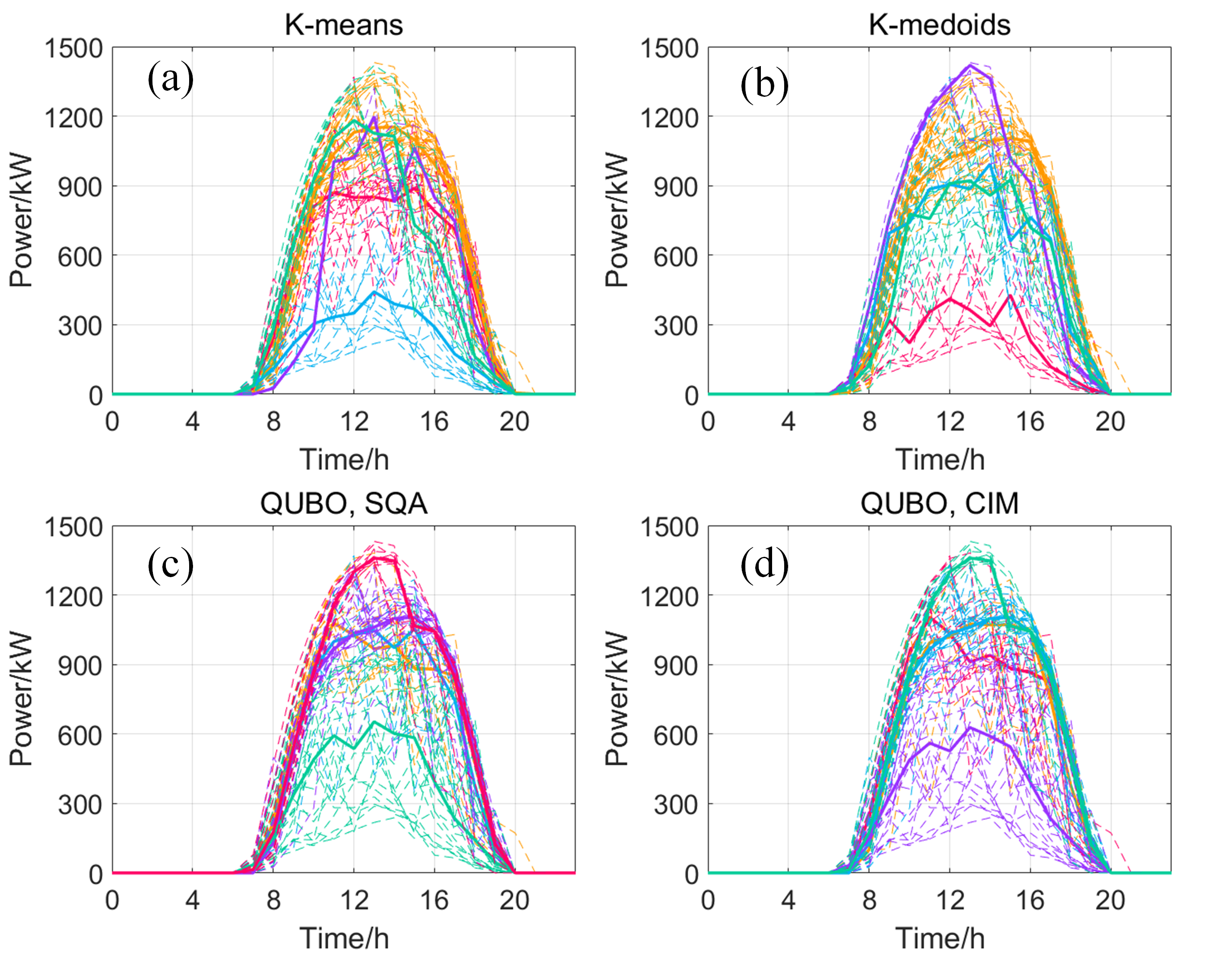}
    \vspace{-0.4cm}
    \caption{Plots for $80$ PV profiles clustered into $5$ groups using different algorithms: $\text{(a)}$ K-means; $\text{(b)}$ K-medoids; $\text{(c)}$ kernel-based QUBO model solved by SQA; $\text{(d)}$ kernel-based QUBO model solved by CIM.}
    \label{fig4}
\end{figure}

\subsubsection{Computational Performance} 
The computing efficiency of these methods is tested to demonstrate the ability of CIM to overcome CoD as shown in Fig.~\ref{fig3}. $\text{(a)}$.
It is clearly shown that the computing time for the CIM has negligible variations with the increase of the input variable while all the other methods computed in the classic computer show a significant increase in the computing time. This apparently demonstrates the CoD that the classical computer suffers when facing an NP-hard clustering problem, while the CIM is barely influenced. Here, the computing time of the CIM fluctuates around $3$ms, reaching a maximum of $3.105$ms in all the test cases in this paper, as shown by the green line in Fig.~\ref{fig3}. $\text{(a)}$. This makes sense since the fiber cavity length in the CIM provided by QBoson is constant \cite{bose} and requires the photon pulse to complete a lap of approximately $3$ms, which matches
the results in Fig.~\ref{fig3}. $\text{(a)}$.
It should be noted that the SQA method uses classical bits to simulate quits, which leads to significantly longer computing times compared to other methods and makes it more susceptible to CoD.
For all Gurobi experiments, a time limit of 300 seconds is imposed to ensure consistency across the test cases. Although feasible solutions are obtained for the 100 and 180 variable cases, the solver failed to return valid results for the 280 and 400 variable cases within the time limit. The corresponding optimality gaps are 84.59\% and 127.19\%, and the solutions do not satisfy the clustering constraints, resulting in silhouette coefficients of zero.
In contrast, the CIM produces valid clustering results with zero optimality gap across all test cases, matching the global optima found by Gurobi and SQA.
Although most methods complete quickly on the current test scale, differences in growth trends become critical in larger-scale applications, where CIM does not suffer from such a CoD.

\subsubsection{Clustering Effectiveness} Silhouette coefficients under different clustering algorithms and variable scales are shown in Fig.~\ref{fig3}. $\text{(b)}$. Overall, the silhouette coefficient of each algorithm tends to decrease as the number of groups increases. 
When comparing the silhouette coefficients of different algorithms in each case, the method proposed in this paper performs similarly to K-medoids and K-means. 
The slight difference in silhouette coefficients is due to the fact that traditional methods directly optimize Euclidean clustering metrics, while our method aims to achieve intra-group similarity in a high-dimensional kernel space. Despite this, the proposed approach achieves a near-constant computing time, making it a practical alternative for large-scale, high-dimensional clustering tasks. As a trade-off, a minor sacrifice in accuracy can be acceptable.
This further demonstrates the applicability of the proposed method.
The results of the clustering of $80$ PV profiles into $5$ groups are then shown in Fig.~\ref{fig4}. The method proposed in this paper achieves more compact intra-group distances compared to the K-means and K-medoids, although it shows a slight decrease in inter-group separation.

\begin{figure}
    \centering
    \includegraphics[scale=1]{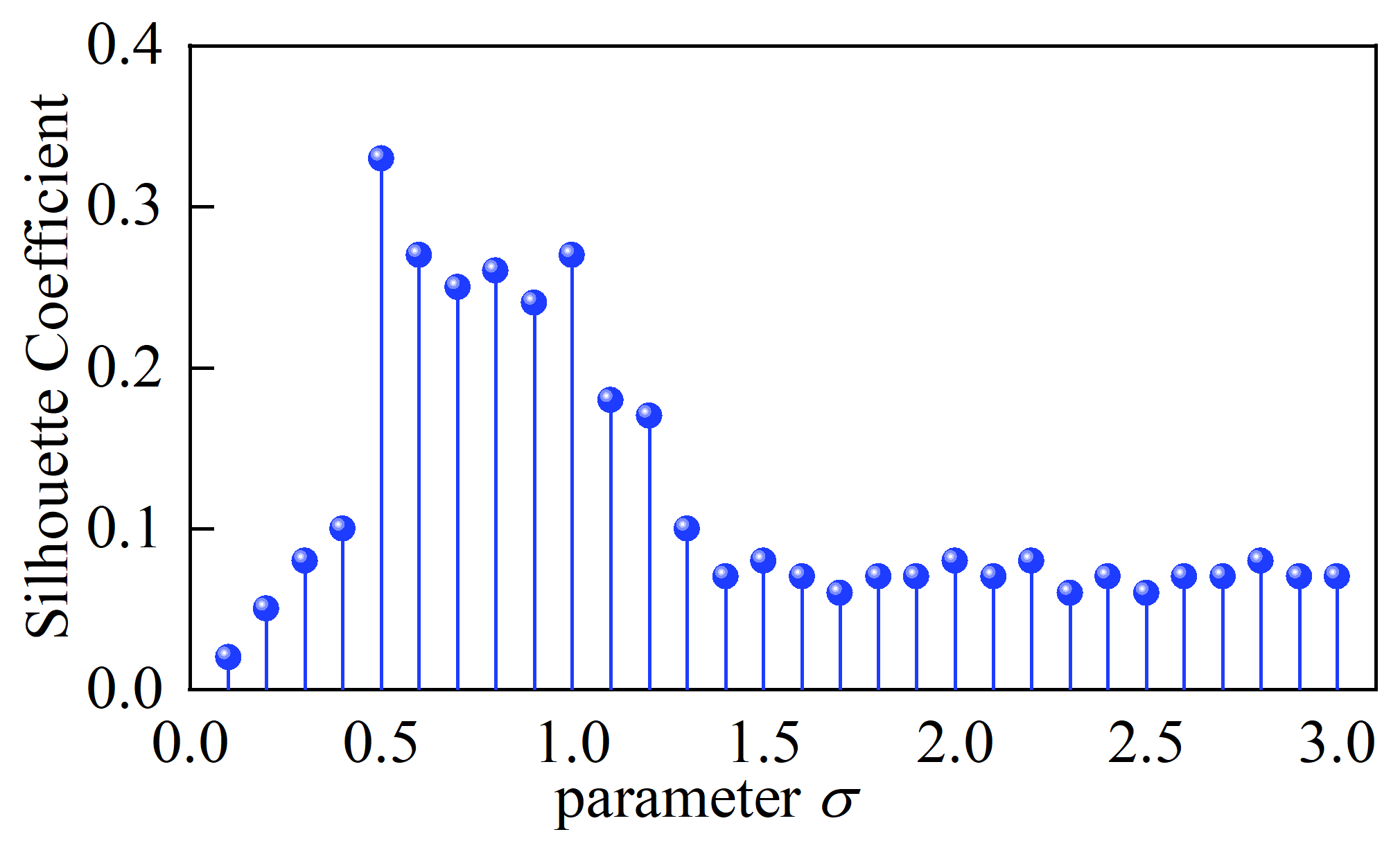}
    \vspace{-0.4cm}
    \caption{Plot for sensitivity of clustering performance to kernel parameter $\sigma$ in the 400 variable case.}
    \label{fig5}
\end{figure}

We further tested the sensitivity of the proposed method to the kernel parameter $\sigma$ in the 400-variable case, as shown in Fig.~\ref{fig5}. The silhouette coefficient peaks around $\sigma=0.5$ and decreases when $\sigma$ is too small or too large. This indicates that the method performs reliably within a reasonable parameter range.

\subsection{Discussion}
\subsubsection{Discussion on Scalability}Although the number of spins limits current CIM hardware, quantum systems are rapidly evolving, with larger-scale devices expected soon \cite{morstyn2024opportunities}. Our QUBO-based modeling framework is inherently scalable once sufficient qubits are accessible, making it compatible with future quantum computers.
To address current limitations, hybrid classical-quantum strategies such as Benders decomposition \cite{FU2023129314}, ADMM \cite{fu2025}, and subQUBO \cite{atobe2021hybrid} can partition large problems into smaller subQUBOs solvable with existing hardware. These methods improve scalability by reducing the QUBO model size executed in the quantum computer.
We will also explore machine learning-assisted approaches to enhance scalability in the future.

\subsubsection{Discussion on Application Scenarios and Advantages}
While the clustering quality of CIM is comparable to classical methods, its advantage lies in specific application scenarios. In real-time power system optimization problems, such as dynamic load aggregation and fast response control, the millisecond-level solving speed of CIM is critical \cite{morstyn2024opportunities}. Moreover, the QUBO-based kernel clustering method benefits from the ability of kernel functions to project data into a high-dimensional feature space, allowing it to effectively handle non-convex datasets where traditional distance-based methods like K-means may fail \cite{camastra2005novel}. These features make the method particularly suitable for time-critical and complex clustering tasks in power systems.

In summary, these results highlight the viability of the proposed method for efficient and accurate clustering in large-scale quantum-inspired applications.

\section{Conclusion}

This paper proposed a novel method for clustering renewable energy profiles, which uses quantum computing technology combined with kernel clustering. From the results of the real quantum computer test, we can conclude that the proposed method provides a good combination of accuracy and computation speed. 
The QUBO-based methodology is also extensible to other power system applications involving discrete optimization, such as grid restoration \cite{FU2023129314} unit commitment\cite{10381746}, \cite{fu2025}.
More importantly, it shows great potential to alleviate CoD compared to other traditional clustering methods such as K-means and K-medoid. 
Clustering of renewable energy profiles has many application scenarios in the field of power systems. In the future, we plan to test its performance in some explicit application scenarios, e.g., unit commitment problems considering renewable energy resources. Additionally, hybrid classical-quantum algorithms will be explored to enhance scalability of the proposed method.

\begin{appendices}
\section{Proof of Isomorphism of Ising and QUBO models}
\renewcommand{\theequation}{A.\arabic{equation}}
\setcounter{equation}{0}
Since, $s_{i} = 2x_{i} - 1$, we have
\begin{equation}
\begin{split}
H_\text{Ising} & = \underset{i \textless j}{\sum} 4 J_{i , j} x_{i} x_{j} - \underset{i}{\sum} x_{i} \underset{j \textgreater i}{\sum} 2 J_{i , j} \\
- & \underset{i \textless j}{\sum} 2 J_{i , j} x_{j} + \underset{i}{\sum} 2 h_{i} x_{i} + C_{1}.
\end{split}
\end{equation}
where $C_1$ is a new constant different from $C$.
Let $a_{i} = - \sum_{j \textgreater i} 2 J_{i , j} + 2 h_{i}$. Since the constant term does not affect the optimization result, $C_1$ is omitted, then
\begin{equation}
\label{Isingmiddle}
H_\text{Ising} = \underset{i \textless j}{\sum} 4 J_{i , j} x_{i} x_{j} + \underset{i}{\sum} a_{i} x_{i} - 2 \underset{i \textless j}{\sum} J_{i , j} x_{j}.
\end{equation}

Now, let us expand the last term to obtain
\begin{equation}
\begin{split}
& \underset{i \textless j}{\sum}J_{i , j} x_{j} = J_{1, 2} x_{2}  + (J_{1, 3} x_{3} + J_{2, 3} x_{3}) + \ldots \\
& + (J_{1, N} x_{N} + \ldots + J_{N-1, N} x_{N}) = \underset{i}{\sum} b_{i} x_{i}.
\end{split}
\end{equation}

Thus,  \eqref{Isingmiddle} is reformulated as
\begin{equation}
\begin{split}
H_\text{Ising} & = \underset{i \textless j}{\sum} 4 J_{i , j} x_{i} x_{j} + \underset{i}{\sum} \left(a_{i}-2b_{i}\right) x_{i}.
\end{split}
\end{equation}

Let $q_{i,j}=2J_{i,j}$ and $q_{i}=a_{i}-2b_{i}$, which yields 
\begin{equation}
H_\text{Ising}= \underset{i}{\sum} \underset{j}{\sum} q_{i,j} x_{i} x_{j} + \underset{i}{\sum} q_{i} x_{i}.
\end{equation} 

Using $C=0$, we could represent an Ising model, \eqref{Ising}, in the QUBO form of \eqref{fqubo}.

\vspace{-0.2cm}
\section{Theoretical Analysis and Complexity Comparison}
The CIM operate on the basis of the physics of DOPO networks. As described in our previous work~\cite{PRA}, the total time required for the CIM to reach a solution is primarily determined by the round-trip time of the photon pulses in the optical cavity. This time is fixed by the cavity length and remains nearly constant, regardless of the number of spins or the size of the problem. This makes the CIM fundamentally different from classical solvers, whose computing time typically grows with the number of variables. The network's evolution is governed by the minimum gain principle, which drives the system toward configurations that minimize photon loss, corresponding to low-energy solutions of the associated Ising Hamiltonian. To provide a clearer comparison, Table~\ref{tab:complexity} summarizes the asymptotic time complexity of several methods applied to clustering problems.

\begin{table}[H]
\centering
\caption{Theoretical time complexity comparison of representative clustering method}
\label{tab:complexity}
\begin{tabular}{c c c}
\toprule
\textbf{Method} & \textbf{Time Complexity} & \textbf{Optimality} \\
\midrule
K-means & $\mathcal{O}(NG)$ & local \\
K-medoids & $\mathcal{O}(N^{2}G)$ & local \\
SQA & $\mathcal{O}\left((NG)^{2} \sim (NG)^{3}\right)$ & local \\
Gurobi (QUBO) & $\mathcal{O}(2^{NG})$ & global \\
CIM & $\mathcal{O}(1)$ & local \\
\bottomrule
\end{tabular}
\end{table}

While the CIM does not provide a mathematical guarantee of global optimality, its dynamics are guided by physical mechanisms that favor low-energy solutions. The solution selection process relies on the stability of low-photon-loss configurations in the DOPO network, which are closely aligned with the optimal or near-optimal Ising configurations \cite{PRA}.

\end{appendices}

 \newcommand{\BIBdecl}{\setlength{\itemsep}{0.01 em}}
 \bibliographystyle{IEEEtran}
\bibliography{IEEEabrv,References.bib}

\end{document}